\documentclass[10pt,a4paper]{article} 

\usepackage{textcomp}
\usepackage{mathrsfs}
\usepackage[utf8]{inputenc}
\usepackage{lmodern}
\usepackage[T1]{fontenc}
\usepackage{mathrsfs}
\usepackage{amsmath}
\usepackage{amssymb}
\usepackage{bussproofs}
\usepackage{thrmappendix}
\usepackage{color}
\usepackage{url}
\usepackage{stmaryrd}
\usepackage{a4wide}
\usepackage{tikz}
\usetikzlibrary{shapes}
\usetikzlibrary{arrows}

\newcommand{\setone}{X}
\newcommand{\funone}{F}
\newcommand{\RR}{\mathbb{R}}

\newcommand{\idxone}{{i}}
\newcommand{\idxtwo}{{h}}

\newcommand{\idxfour}{{m}}
\newcommand{\Idxone}{I}
\newcommand{\Idxtwo}{\mathcal{H}}

\newcommand{\distrsp}[2]{\mathscr{D}^{#1}_{#2}}
\newcommand{\distrone}{\mathscr{E}}
\newcommand{\distrtwo}{\mathscr{F}}
\newcommand{\distrthree}{\mathscr{G}}

\newcommand{\supp}[1]{\mathfrak{S}\left({#1}\right)}
\newcommand{\probone}{p}
\newcommand{\probtwo}{q}

\newcommand{\transitionpm}{\mathcal{T}} 
\newcommand{\transpm}[1]{\mathcal{T}_{#1}}
\newcommand{\lmcone}{\mathscr{M}} 
\newcommand{\lmcof}[1]{\mathscr{M}_{#1}} 
\newcommand{\testone}{t}
\newcommand{\testtwo}{s}
\newcommand{\testthree}{p}
\newcommand{\contest}[2]{\langle #1,#2\rangle}

\newcommand{\complexsp}{\mathbb{C}}
\newcommand{\coeffone}{\alpha}
\newcommand{\ket}[1]{\lvert { #1}\rangle}
\newcommand{\newop}[2]{\mathtt{NW}_{#1}^{#2}}
\newcommand{\projop}[2]{\mathtt{MS}_{#1}^{#2}}
\newcommand{\probop}[2]{\mathtt{PR}_{#1}^{#2}}

\newcommand{\qubitone}{b}

\newcommand{\qubitty}{\mathtt{qbit}}
\newcommand{\quclosure}[2]{\left[{#1},{#2}\right]}

\newcommand{\quregket}[6]{\ket{{#1}\leftarrow {#2},{#3}\leftarrow {#4},\cdots,{#5}\leftarrow {#6}}}
\newcommand{\quregone}{\mathscr{Q}}
\newcommand{\quregtwo}{\mathscr{U}}
\newcommand{\quregthree}{\mathscr{W}}
\newcommand{\quvarone}{r}

\newcommand{\quvars}{\mathcal{Q}}
\newcommand{\sbsts}[1]{\mathcal{SB}(#1)}
\newcommand{\sbstone}{\eta}
\newcommand{\hilb}[1]{\mathcal{H}(#1)}
\newcommand{\unopone}{U}

\newcommand{\linL}{\ensuremath{\ell\mathit{ST}_{\lambda}}}	
\newcommand{\linPL}{\ensuremath{\ell\mathit{PST}_{\lambda}}}
\newcommand{\linQL}{\ensuremath{\ell\mathit{QST}_{\lambda}}} 

\newcommand{\bnf}{::=}
\newcommand{\midd}{\; \; \mbox{\Large{$\mid$}}\;\;}

\newcommand{\termone}{e}
\newcommand{\termtwo}{f}
\newcommand{\termthree}{g}
\newcommand{\termfour}{h}
\newcommand{\termfive}{s}
\newcommand{\termsix}{r}
\newcommand{\termseven}{t}
\newcommand{\varone}{x}
\newcommand{\vartwo}{y}
\newcommand{\varthree}{z}
\newcommand{\termtt}{\mathtt{tt}}
\newcommand{\termff}{\mathtt{ff}}
\newcommand{\bitone}{b}
\newcommand{\termif}[3]{\mathtt{if}\;#1\;\mathtt{then}\;#2\;\mathtt{else}\;#3}
\newcommand{\termlet}[4]{\mathtt{let}\;#1\;\mathtt{be}\;\langle #2,#3\rangle\;\mathtt{in}\;#4}
\newcommand{\termpair}[2]{\langle #1,#2\rangle}
\newcommand{\termvector}[1]{\langle #1\rangle}
\newcommand{\termdiv}{\Omega}
\newcommand{\boolvalone}{b}
\newcommand{\terms}[1]{\mathcal{T}^{#1}}
\newcommand{\termsp}[2]{\mathcal{T}^{#1}_{#2}}
\newcommand{\termsc}[1]{\mathcal{E}^{#1}}
\newcommand{\values}[1]{\mathcal{V}^{#1}}
\newcommand{\valuesp}[2]{\mathcal{V}_{#2}^{#1}}
\newcommand{\types}{\mathcal{Y}}
\newcommand{\abstr}[2]{\lambda #1.#2}
\newcommand{\weak}[2]{\mathtt{weak}\;#1\;\mathtt{in}\;#2}
\newcommand{\valone}{v}
\newcommand{\valtwo}{w}
\newcommand{\valthree}{u}
\newcommand{\valfour}{z}
\newcommand{\subst}[3]{#1\{#3/#2\}}
\newcommand{\boolty}{\mathtt{bool}}
\newcommand{\typeone}{A} 
\newcommand{\typetwo}{B}
\newcommand{\typethree}{E} 
\newcommand{\tj}[3]{#1\vdash #2:#3}
\newcommand{\emcon}{\emptyset} 
\newcommand{\contone}{\Gamma}
\newcommand{\conttwo}{\Delta} 

\newcommand{\ssred}{\rightarrow}
\newcommand{\bsred}{\Downarrow}
\newcommand{\termabs}[2]{\lambda\,{#1}.{#2}}
\newcommand{\ctxhole}{[\cdot]}
\newcommand{\ctxone}{C}

\newcommand{\tjc}[5]{#1\vdash #2[#3\vdash #4]:#5}
\newcommand{\conpre}{\leq} 
\newcommand{\conprep}[1]{\leq_{#1}}
\newcommand{\conequ}{\equiv} 
\newcommand{\conequp}[1]{\equiv_{#1}}
\newcommand{\termand}{\mathtt{and}}
\newcommand{\termor}{\mathtt{or}}
\newcommand{\termchoice}[2]{{#1}\oplus{#2}}
\newcommand{\semantic}[1]{\mbox{\textlbrackdbl} {#1}
  \mbox{\textrbrackdbl}}
\newcommand{\econe}{E} 
\newcommand{\termnew}{\mathtt{new}}
\newcommand{\termmeas}{\mathtt{meas}}
\newcommand{\termU}{U}
 
\newcommand{\states}{\mathcal{S}} 
\newcommand{\statone}{s}
\newcommand{\stattwo}{t}

\newcommand{\labels}{\mathcal{L}}
\newcommand{\transitionrel}{\mathscr{T}} 
\newcommand{\ltsone}{\mathscr{L}}
\newcommand{\labelone}{\ell}
\newcommand{\ltsof}[1]{\mathscr{L}_{#1}}
\newcommand{\eval}{\mathit{eval}}
\newcommand{\trans}[1]{\transitionrel_{#1}}
\newcommand{\dist}[1]{\widehat{#1}}
\newcommand{\relone}{R}

\newcommand{\appsim}{\preceq}
\newcommand{\appsimp}[1]{\preceq_{#1}}
\newcommand{\appbis}{\sim}
\newcommand{\appbisp}[1]{\sim_{#1}}
\newcommand{\howe}[1]{{#1}^H}

\newenvironment{varitemize}
{
\begin{list}{\labelitemii}
{
\setlength{\itemsep}{0pt}
 \setlength{\topsep}{0pt}
 \setlength{\parsep}{0pt}
 \setlength{\partopsep}{0pt}
 \setlength{\leftmargin}{15pt}
 \setlength{\rightmargin}{0pt}
 \setlength{\itemindent}{0pt}
 \setlength{\labelsep}{5pt}
 \setlength{\labelwidth}{10pt}}}
{
 \end{list}
}

\newcounter{number}

\newenvironment{varenumerate}
{
\begin{list}{\arabic{number}.}
{
  \usecounter{number}
  \setlength{\itemsep}{0pt}
  \setlength{\topsep}{0pt}
  \setlength{\parsep}{0pt}
  \setlength{\partopsep}{0pt}
  \setlength{\leftmargin}{15pt}
  \setlength{\rightmargin}{0pt}
  \setlength{\itemindent}{0pt}
  \setlength{\labelsep}{5pt}
  \setlength{\labelwidth}{15pt}
}}
{
\end{list} 
}

\newtheorem{theorem}{Theorem}
\newtheorem{lemma}{Lemma}

\newtheorem{example}{Example}

\title{Applicative Bisimulation\\ and Quantum $\lambda$-Calculi\\ (Long Version)\thanks{This work is partially supported by the
  ANR project 12IS02001 PACE.}}

\author{Ugo Dal Lago \and Alessandro Rioli}

\begin{document}

\maketitle

\begin{abstract}
  Applicative bisimulation is a coinductive technique to check program
  equivalence in higher-order functional languages. It is known to be
  sound --- and sometimes complete --- with respect to context
  equivalence. In this paper we show that applicative bisimulation
  also works when the underlying language of programs takes the form
  of a \emph{linear} $\lambda$-calculus extended with features such as
  probabilistic binary choice, but also quantum data, the latter being
  a setting in which linearity plays a role. The main results are
  proofs of soundness for the obtained notions of bisimilarity.
\end{abstract}

\section{Introduction}
Program equivalence is one of the fundamental notions in the theory of
programming languages. Studying the nature of program equivalence is
not only interesting from a purely foundational point of view, but can
also be the first step towards defining (semi)automatic techniques for
program verification, or for validating compiler optimisations.
As an example, conformance of a program to a specification often corresponds
to the equivalence between the program and the specification, once the
latter is written in the same formalism as the program.

If the language at hand is an higher-order functional language,
equivalence is traditionally formalised as Morris' \emph{context
  equivalence}: two programs are considered equivalent if and only if
they have the same behavior in \emph{every possible}
context~\cite{Morris-68}. This makes it relatively easy to prove two
programs to be \emph{not} equivalent, since this merely amounts to
finding \emph{one} context which separates them. On the other hand,
proving two terms to be equivalent requires one to examine their
behaviour in \emph{every possible} context.

Various ways to alleviate the burden of proving context equivalence
have been proposed in the literature, from CIU theorems (in which the
class of contexts is restricted without altering the underlying
relation~\cite{Milner-77}) to adequate denotational semantics, to
logical relations~\cite{Plotkin-73}. We are here interested in
coinductive techniques akin to bisimulation. Indeed, they have been
shown to be very powerful, to the point of not only being sound, but
even complete as ways to prove terms to be context
equivalent~\cite{Pitts-97}. Among the various notions of bisimulation
which are known to be amenable to higher-order programs, the simplest
one is certainly Abramsky's applicative
bisimulation~\cite{Abramsky-90}, in which terms are seen as
interactive objects and the interaction with their environment
consists in taking input arguments or outputting observable results.

Applicative bisimulation is indeed well-known to be fully-abstract
w.r.t. context equivalence when instantiated on plain, untyped,
\emph{deterministic} $\lambda$-calculi~\cite{Abramsky-90}. When the
calculus at hand also includes a choice operator, the situation is
more complicated: while applicative bisimilarity is invariably a
congruence, thus sound for context equivalence, completeness generally
fails ~\cite{Pitts-97,Lassen-97}, even if some unexpected positive
results have recently been obtained by Crubill\'e and the first
author~\cite{DalLagoCrubille-14} in a probabilistic setting. An
orthogonal issue is the one of linearity: does applicative
bisimulation work well when the underlying calculus has linear types?
The question has been answered positively, but only for deterministic
$\lambda$-calculi~\cite{Crole-01,Bierman00}.  Finally, soundness does
not hold in general if the programming language at hand has
references~\cite{KoutavasLevySumii-11}.

In this paper, we define and study applicative bisimulation when
instantiated on linear $\lambda$-calculi, starting with a purely
deterministic language, and progressively extending it with
probabilistic choice and quantum data, a setting in which linearity is
an essential ingredient~\cite{SelingerValiron-05,SelingerValiron-08}.
The newly added features in the language are shown to correspond to
mild variations in the underlying transition system, which in presence
of probabilistic choice becomes a labelled Markov chain. The main
contributions of this paper are congruence results for applicative
bisimilarity in probabilistic and quantum $\lambda$-calculi, with
soundness with respect to context equivalence as an easy corollary. In
all the considered calculi, Howe's technique~\cite{Howe-96,Pitts-97}
plays a key role.

This is the first successful attempt to apply coinductive techniques
to quantum, higher-order, calculi. The literature offers some ideas
and results about bisimulation and simulation in the context of
quantum process
algebras~\cite{GayNagarajan-05,DengFeng-12,DavidsonGayMlnarikNagarajanPapanikolaou-12}.
Deep relations between quantum computation and coalgebras have
recently been discovered~\cite{Jacobs-11}. None of the cited works,
however, deals with higher-order functions.

This paper is structured as follows. In Section~\ref{sect:linear}, a
simple linear $\lambda$-calculus, called $\linL$ will be introduced,
together with its operational semantics. This is a purely
deterministic calculus, on top of which our extensions will be
defined. Section~\ref{sect:appbis} presents the basics of applicative
bisimulation, instantiated on $\linL$. A probabilistic variation on
$\linL$, called $\linPL$, is the subject of
Section~\ref{sect:probchoice}, which also discusses the impact of
probabilities to equivalences and bisimilarity.
Section~\ref{sect:quantum} is about a quantum variation on $\linL$,
dubbed $\linQL$, together with a study of bimilarity for it.
Section~\ref{sect:fullabst} concludes the paper with a discussion
about full-abstraction. An extended version of this paper with more
details is available~\cite{EV}. 
\section{Linear $\lambda$-Calculi: A Minimal Core}\label{sect:linear}
In this section, a simple linear $\lambda$-calculus called $\linL$
will be introduced, together with the basics of its operational
semantics. \emph{Terms} and \emph{values} are generated by the
following grammar:
\begin{align*}
\termone,\termtwo&\bnf\valone\midd\termone\termone\midd
  \termif{\termone}{\termone}{\termone}\midd\termlet{\termone}{\varone}{\varone}{\termone}\midd\termdiv;\\
\valone,\valtwo&\bnf\varone\midd\termtt\midd\termff\midd\lambda\varone.\termone\midd\termpair{\valone}{\valone}.
\end{align*}
Observe the presence not only of abstractions and applications, but
also of value pairs, and of basic constructions for booleans. Pairs
of arbitrary terms can be formed as follows, as syntactic sugar:
$$
\termpair{\termone}{\termtwo}=(\lambda\varone.\lambda\vartwo.\termpair{\varone}{\vartwo})\termone\termtwo.
$$
Finally, terms include a constant $\termdiv$ for divergence. $\bitone$
is a metavariable for truth values, i.e. $\bitone$ stands for either
$\termtt$ or $\termff$.  We need a way to enforce linearity, i.e., the
fact that functions use their arguments \emph{exactly} once. This can
take the form of a linear type system whose language of \emph{types}
is the following: 
$$
\typeone,\typetwo\bnf\boolty\midd\typeone\multimap\typeone\midd\typeone\otimes\typeone.
$$ 
The set $\types$ includes all types. \emph{Typing judgments} are in the form
$\tj{\contone}{\termone}{\typeone}$, where $\contone$ is a set of
assignments of types to variables. Typing rules are standard, and can
be found in Figure~\ref{fig:dettyping}.
\begin{figure}
\begin{center}
\fbox{
\begin{minipage}{.97\textwidth}
  $$
  \AxiomC{}
  \UnaryInfC{$\tj{\varone:\typeone}{\varone}{\typeone}$} 
  \DisplayProof
  \quad
  \AxiomC{}
  \UnaryInfC{$\tj{}{\bitone}{\boolty}$} 
  \DisplayProof
  \quad
  \AxiomC{$\tj{\contone}{\termone}{\typeone\multimap\typetwo}$} 
  \AxiomC{$\tj{\conttwo}{\termtwo}{\typeone}$} 
  \BinaryInfC{$\tj{\contone,\conttwo}{\termone\termtwo}{\typetwo}$}    
  \DisplayProof
  \quad 
  \AxiomC{}
  \UnaryInfC{$\tj{\contone}{\termdiv}{\typeone}$}
  \DisplayProof
  $$
  $$
  \AxiomC{$\tj{\contone}{\valone}{\typeone}$}
  \AxiomC{$\tj{\conttwo}{\valtwo}{\typetwo}$}
  \BinaryInfC{$\tj{\contone,\conttwo}{\termpair{\valone}{\valtwo}}{\typeone\otimes\typetwo}$}    
  \DisplayProof
  \qquad
  \AxiomC{$\tj{\contone,x:X,y:Y}{\termone}{\typeone}$} 
  \AxiomC{$\tj{\conttwo}{\termtwo}{X\otimes Y}$}
  \BinaryInfC{ $\tj{\contone,\conttwo}{\termlet{\termtwo}{x}{y}{\termone}}{\typeone} $ }
  \DisplayProof
  $$
  $$
  \AxiomC{$\tj{\contone,\varone:\typeone}{\termone}{\typetwo}$}
  \UnaryInfC{$\tj{\contone}{\lambda\varone.\termone}{\typeone\multimap\typetwo}$}
  \DisplayProof
  \qquad
  \AxiomC{$\tj{\contone}{\termone}{\boolty}$}
  \AxiomC{$\tj{\conttwo}{\termtwo}{\typeone}$}
  \AxiomC{$\tj{\conttwo}{\termthree}{\typeone}$}
  \TrinaryInfC{$\tj{\contone,\conttwo}{\termif{\termone}{\termtwo}{\termthree}}{\typeone}$}
  \DisplayProof
  $$
  \vspace{3pt}
\end{minipage}}
\caption{Typing Rules}\label{fig:dettyping}
\end{center}
\end{figure}
The set $\termsp{\linL}{\contone,\typeone}$ contains all terms
$\termone$ such that
$\tj{\contone}{\termone}{\typeone}$. $\termsp{\linL}{\emcon,\typeone}$
is usually written as $\termsp{\linL}{\typeone}$. Notations like
$\valuesp{\linL}{\contone,\typeone}$ or $\valuesp{\linL}{\typeone}$
are the analogues for values of the corresponding notations for terms.

Endowing $\linL$ with call-by-value small-step or big-step semantics
poses no significant problem. In the first case, one defines a binary
relation $\ssred$ between closed terms of any type by the usual rule
for $\beta$-reduction, the natural rule for the conditional operator,
and the following rule:
$\termlet{\termpair{\valone}{\valtwo}}{\varone}{\vartwo}{\termone}\ssred\subst{\termone}{\varone,\vartwo}{\valone,\valtwo}$.
Similarly, one can define a big-step evaluation relation $\bsred$
between closed terms and values by a completely standard set of rules
(see \cite{EV} for more details). The expression $\termone\bsred$, as
usual, indicates the existence of a value $\valone$ with
$\termone\bsred\valone$.  Subject reduction holds in the following
sense: if $\tj{\emcon}{\termone}{\typeone}$, $\termone\ssred\termtwo$,
and $\termone\bsred\valone$, then both
$\tj{\emcon}{\termtwo}{\typeone}$ and
$\tj{\emcon}{\valone}{\typeone}$.

The expressive power of the just-introduced calculus is rather
poor. Nonetheless, it can be proved to be complete for first-order
computation over booleans, in the following sense: for every function
$\funone:\{\termtt,\termff\}^n\rightarrow\{\termtt,\termff\}$, there
is a term which \emph{computes} $\funone$, i.e. a term
$\termone_\funone$ such that
$\termone_\funone\termvector{\boolvalone_1,\ldots,\boolvalone_n}\bsred\funone(\boolvalone_1,\ldots,\boolvalone_n)$
for every
$\boolvalone_1,\ldots,\boolvalone_n\in\{\termtt,\termff\}^n$. Indeed,
even if copying and erasing bits is not in principle allowed, one
could anyway encode, e.g., duplication as the following combinator of
type $\boolty\multimap\boolty\otimes\boolty$:
$\lambda\varone.\termif{\varone}{\termpair{\termtt}{\termtt}}{\termpair{\termff}{\termff}}$.
Similarly, if $\tj{\contone}{\termone}{\typeone}$ and $\varone$ is a fresh
variable, one can easily find a term $\weak{\varone}{\termone}$ such that
$\tj{\contone,\varone:\boolty}{\weak{\varone}{\termone}}{\typeone}$ and 
$\weak{\bitone}{\termone}$ behaves like $\termone$ for every $\bitone\in\{\termff,\termtt\}$.

But how could one capture program equivalence in an higher-order
setting like the one we are examining? The canonical answer goes back
to Morris~\cite{Morris-68}, who proposed \emph{context} equivalence
(also known as \emph{observational} equivalence) as the right way to
compare terms. Roughly, two terms are context equivalent iff they
behave the same when observed in any possible \emph{context},
i.e. when tested against any possible \emph{observer}. Formally, a
context is nothing more than a term with a single occurrence of a
special marker called the \emph{hole} and denoted as $\ctxhole$
(see~\cite{EV}). Given a context $\ctxone$ and a term $\termone$,
$\ctxone[\termone]$ is the term obtained by filling the single
occurrence of $\ctxhole$ in $\ctxone$ with $\termone$. For contexts to
make sense in a typed setting, one needs to extend typing rules to
contexts, introducing a set of rules deriving judgments in the form
$\tjc{\contone}{\ctxone}{\conttwo}{\typeone}{\typetwo}$, which can be
read informally as saying that whenever
$\tj{\conttwo}{\termone}{\typeone}$, it holds that
$\tj{\contone}{\ctxone[\termone]}{\typetwo}$.

We are now in a position to define the context preorder: given two
terms $\termone$ and $\termtwo$ such that
$\tj{\contone}{\termone,\termtwo}{\typeone}$, we write
$\termone\conprep{\contone,\typeone}\termtwo$ iff for every context
$\ctxone$ such that
$\tjc{\emcon}{\ctxone}{\contone}{\typeone}{\typetwo}$, if
$\ctxone[\termone]\bsred$ then $\ctxone[\termtwo]\bsred$. If
$\termone\conprep{\contone,\typeone}\termtwo$ and
$\termtwo\conprep{\contone,\typeone}\termone$, then $\termone$ and
$\termtwo$ are said to be \emph{context equivalent}, and we write
$\termone\conequp{\contone,\typeone}\termtwo$. What we have just
defined, infact, are two \emph{typed relations} $\conpre$ and
$\conequ$, that is to say two families of relations indexed by
contexts and types, i.e. $\conpre$ is the family
$\{\conprep{\contone,\typeone}\}_{\contone,\typeone}$, while $\conequ$
is $\{\conequp{\contone,\typeone}\}_{\contone,\typeone}$.  If in the
scheme above the type $\typetwo$ is restricted so as to be $\boolty$,
then the obtained relations are the \emph{ground} context preorder and
\emph{ground} context equivalence, respectively. Context equivalence
is, almost by construction, a congruence. Similarly, the context
preorder is easily seen to be a precongruence.

\section{Applicative Bisimilarity and its Properties}\label{sect:appbis}
Context equivalence is universally accepted as the canonical notion of
equivalence of higher-order programs, being robust, and only relying
on the underlying operational semantics. Proving terms \emph{not}
context equivalent is relatively easy: ending up with a single context
separating the two terms suffices. On the other hand, the universal
quantification over all contexts makes proofs of equivalence hard.

A variety of techniques have been proposed to overcome this problem,
among them logical relations, adequate denotational models and context
lemmas.  As first proposed by Abramsky~\cite{Abramsky-90}, coinductive
methodologies (and the bisimulation proof method in particular) can be
fruitfully employed.  Abramsky's \emph{applicative} bisimulation is
based on taking argument passing as the basic interaction mechanism:
what the environment can do with a $\lambda$-term is either evaluating
it or passing it an argument.

In this section, we will briefly delineate how to define applicative
bisimilarity for the linear $\lambda$-calculus $\linL$. We will do
that in an unnecessarily pedantic way, defining a labelled transition
system, and then playing the usual bisimulation game on top of
it. This has the advantage of making the extensions to probabilistic
and quantum calculi much easier.

A \emph{labelled transition system} (LTS in the following) is a triple
$\ltsone=(\states,\labels,\transitionrel)$, where $\states$ is a set
of \emph{states}, $\labels$ is a set of \emph{labels}, and
$\transitionrel$ is a subset of $\states\times\labels\times\states$.
If for every $\statone\in\states$ and for every $\labelone\in\labels$
there is \emph{at most} one state $\stattwo\in\states$ with
$(\statone,\labelone,\stattwo)\in\transitionrel$, then $\ltsone$ is
said to be \emph{deterministic}. The theory of bisimulation for LTSs
is very well-studied~\cite{SangiorgiBook-12} and forms one of the
cornerstones of concurrency theory.

An applicative bisimulation relation is nothing more than a
bisimulation on an LTS $\ltsof{\linL}$ defined \emph{on top of} the
$\lambda$-calculus $\linL$. More specifically, the LTS $\ltsof{\linL}$
is defined as the triple
$$
(\overline{\terms{\linL}}\uplus\overline{\values{\linL}},
 \overline{\termsc{\linL}}\uplus\overline{\values{\linL}}\cup\{\eval,\termtt,\termff\}\cup(\types\uplus\types),\trans{\linL}),
$$
where:
\begin{varitemize}
\item
  $\overline{\terms{\linL}}$ is the set
  $\cup_{\typeone\in\types}(\termsp{\linL}{\typeone}\times\{\typeone\})$, 
  similarly for $\overline{\values{\linL}}$. On the other hand,
  $\overline{\termsc{\linL}}$ is
  $\cup_{\typeone,\typetwo,\typethree\in\types}(\termsp{\linL}{\varone:\typeone,\vartwo:\typetwo,\typethree}\times\{(\typeone,\typetwo,\typethree)\})$.
  Observe how any pair $(\valone,\typeone)$ appears
  twice as a state, once as an element of $\overline{\terms{\linL}}$
  and again as an element of $\overline{\values{\linL}}$. Whenever
  necessary to avoid ambiguity, the 
  second instance will be denoted as $(\dist{\valone},\typeone)$. Similarly
  for the two copies of any type $\typeone$ one finds as labels.
\item
  The label $\eval$ models evaluation of terms, while the labels
  $\termtt,\termff$ are the way a boolean constant declares its own
  value.
\item
  The relation $\trans{\linL}$ contains all triples in the following
  forms:
  $$
  \begin{array}{c}
  ((\dist{\termtt},\boolty), \termtt, (\dist{\termtt},\boolty));
  \qquad
  ((\dist{\termff},\boolty), \termff, (\dist{\termff},\boolty));\\
  ((\dist{\lambda\varone.\termone},\typeone\multimap\typetwo),(\valone,\typeone),(\subst{\termone}{\varone}{\valone},\typetwo));\\
  ((\dist{\termpair{\valone}{\valtwo}},\typeone\otimes\typetwo),(\termone,(\typeone,\typetwo,\typethree)),(\termone\{\valone/\varone,\valtwo/\vartwo\},\typethree));\\
  ((\termone,\typeone),\typeone,(\termone,\typeone));
  \qquad
  ((\dist{\valone},\typeone),\dist{\typeone},(\dist{\valone},\typeone));
  \qquad
  ((\termone, \typeone),\eval, (\dist{\valone},\typeone));
  \end{array}
  $$
  where, in the last item, we of course assume that $\termone\Downarrow\valone$.
\end{varitemize}
Basically, values interact with their environment based on their
types: abstractions take an input argument, pairs gives their two
components to a term which can handle them, and booleans constants
simply expose their value. The only way to interact with terms is by
evaluating them. Both terms and values expose their type.  As one can
easily verify, the labelled transition system $\ltsof{\linL}$ is
deterministic.  Simulation and bisimulation relations for
$\ltsof{\linL}$ are defined as for any other LTS.  Notice, however,
that both are binary relations on \emph{states}, i.e., on elements of
$\overline{\terms{\linL}}\uplus\overline{\values{\linL}}$. Let us
observe that:
\begin{varitemize}
\item
  Two pairs $(\termone,\typeone)$ and $(\termtwo,\typetwo)$ can be put
  in relation only if $\typeone=\typetwo$, because each state makes
  its type public through a label. For similar reasons, states in the
  form $(\valone,\typeone)$ and $(\dist{\valtwo},\typetwo)$ cannot be
  in relation, not even if $\typeone=\typetwo$.
\item
  If $(\valone,\typeone)$ and $(\valtwo,\typeone)$ are in relation,
  then also $(\dist{\valone},\typeone)$ and
  $(\dist{\valtwo},\typeone)$ are in relation. Conversely, if
  $(\dist{\valone},\typeone)$ and $(\dist{\valtwo},\typeone)$ are in a
  (bi)simulation relation $\relone$, then
  $\relone\cup\{((\valone,\typeone),(\valtwo,\typeone))\}$ is itself a
  (bi)simulation.
\end{varitemize}
As a consequence, (bi)similarity can be seen as a relation on terms, indexed
by types. Similarity is denoted as $\appsim$, and its restriction to (closed)
terms of type $\typeone$ is indicated with $\appsimp{\typeone}$. For bisimilarity,
symbols are $\appbis$ and $\appbisp{\typeone}$, respectively. (Bi)similarity
can be generalised to a typed relation by the usual open extension.
\begin{example}
An example of two distinct programs which can be proved bisimilar are the
following:
$$
\termone=\lambda\varone.\lambda\vartwo.\lambda\varthree.\termand\;(\varone\vartwo)\;(\termor\;\varthree\:\termtt);
\qquad
\termtwo=\lambda\varone.\lambda\vartwo.\lambda\varthree.\varone(\termor\;(\termand\;\varthree\;\termff)\;\vartwo);
$$ where $\termand$ and $\termor$ are combinators computing the
eponymous boolean functions. Both $\termone$ and $\termtwo$ can be
given the type
$(\boolty\multimap\boolty)\multimap\boolty\multimap\boolty\multimap\boolty$
in the empty context. They can be proved bisimilar by just giving a
relation $\relone_{e,f}$ which contains the pair $(e,f)$ and which can
be proved to be an applicative bisimulation. Another interesting
example of terms which can be proved bisimilar are the term
$\termone=\termif{\termtwo}{\termthree}{\termfour}$ and the term
$\termfive$ obtained from $\termone$ by $\lambda$-abstracting all
variables which occur free in $\termthree$ (and, equivalently, in
$\termfour$), then applying the same variables to the obtained
term. For more details, see~\cite{EV}.
\end{example}

Is bisimilarity sound for (i.e., included in) context equivalence?
And how about the reverse inclusion? For a linear, deterministic
$\lambda$-calculus like the one we are describing, both questions have
already been given a positive answer \cite{DengFeng-12}. In the next
two sections, we will briefly sketch how the correspondence can be
proved.

\subsection{(Bi)similarity is a (Pre)congruence.}\label{subsect:appbis-congruence}
A natural way to prove that similarity is included in the context preorder, (and
thus that bisimilarity is included in context equivalence) consists in first showing
that similarity is a \emph{precongruence}, that is to say a preorder relation which
is compatible with all the operators of the language.

While proving that $\appsim$ is a preorder is relatively easy, the
naive proof of compatibility (i.e. the obvious induction) fails, due
to application. A nice way out is due to Howe~\cite{Howe-96}, who
proposed a powerful and reasonably robust proof based on so-called
precongruence candidates. Intuitively, the structure of Howe's method
is the following:
\begin{varenumerate}
\item\label{point:howefirst} 
  First of all, one defines an operator $\howe{(\cdot)}$ on typed
  relations, in such a way that whenever a typed relation $\relone$ is
  a preorder, $\howe{\relone}$ is a precongruence.
\item\label{point:howesecond}
  One then proves, again under the condition that $\relone$ is an
  equivalence relation, that $\relone$ is included into
  $\howe{\relone}$, and that $\howe{\relone}$ is substitutive.
\item\label{point:howethird}
  Finally, one proves that $\howe{\appsim}$ is itself an applicative
  simulation. This is the so-called Key Lemma~\cite{Pitts-97},
  definitely the most difficult of the three steps.
\end{varenumerate}
Points \ref{point:howesecond} and \ref{point:howethird} together imply
that $\appsim$ and $\howe{\appsim}$ coincide. But by point
\ref{point:howefirst}, $\howe{\appsim}$, thus also $\appsim$, are
precongruences. Points \ref{point:howefirst} and
\ref{point:howesecond} do not depend on the underlying operational
semantics, but on only on the language's constructs.

In Figure~\ref{fig:howe}, one can find the full set of rules
defining $\howe{(\cdot)}$ when the underlying terms are those of
$\linL$.
\begin{figure}
\begin{center}
\fbox{
\begin{minipage}{.97\textwidth}
  \footnotesize
  \vspace{4pt}
  $$
  \AxiomC{$\tj{\emcon}{c\relone \termseven}{\typeone}$}
  \UnaryInfC{$\tj{\emcon}{c\howe{\relone}\termseven}{\typeone}$} 
  \DisplayProof
  \qquad
  \AxiomC{$\tj{\varone:\typeone}{\varone\relone \termseven}{\typeone}$}
  \UnaryInfC{$\tj{\emcon}{\varone\howe{\relone}\termseven}{\typeone}$} 
  \DisplayProof
  $$
  \vspace{4pt}
  $$
  \AxiomC{$\tj{\contone,\varone:\typetwo}{\termone\howe{\relone} \termfour}{\typeone}$}
  \AxiomC{$\tj{\contone}{(\termabs{\varone}{\termfour})\relone\termseven}{\typetwo\multimap\typeone}$}
  \BinaryInfC{$\tj{\contone}{(\termabs{\varone}{\termone})\howe{\relone}\termseven}{\typetwo\multimap\typeone}$} 
  \DisplayProof
  $$
  \vspace{4pt}
  $$
  \AxiomC{$\tj{\contone}{\termone\howe{\relone} \termfour}{\typetwo\multimap\typeone}$}
  \AxiomC{$\tj{\conttwo}{\termtwo\howe{\relone} \termfive}{\typetwo}$}
  \AxiomC{$\tj{\contone,\conttwo}{(\termfour\termfive)\relone\termseven}{\typeone}$}
  \TrinaryInfC{$\tj{\contone,\conttwo}{(\termone\termtwo)\howe{\relone}\termseven}{\typeone}$} 
  \DisplayProof  
  $$
  \vspace{4pt}
  $$
  \AxiomC{$
    \begin{array}{c}
    \tj{\contone}{\termone\howe{\relone} \termfour}{\boolty}\\
    \tj{\conttwo}{\termtwo\howe{\relone} \termfive}{\typeone} \quad \tj{\conttwo}{\termthree\howe{\relone} \termsix}{\typeone}\\
    \tj{\contone,\conttwo}{(\termif{\termfour}{\termfive}{\termsix})\relone\termseven}{\typeone}\\
    \end{array}$}
  \UnaryInfC{$\tj{\contone,\conttwo}{(\termif{\termone}{\termtwo}{\termthree})\howe{\relone}\termseven}{\typeone}$} 
  \DisplayProof  
  \qquad
  \AxiomC{$
     \begin{array}{c}
       \tj{\contone}{\termone\howe{\relone} \termfour}{X \otimes Y}\\
       \tj{\conttwo,\varone:X,\vartwo:Y}{\termtwo\howe{\relone} \termfive}{\typeone}\\
       \tj{\contone,\conttwo}{(\termlet{\termfour}{x}{y}{\termfive})\relone\termseven}{\typeone}
     \end{array}$}
  \UnaryInfC{$\tj{\contone,\conttwo}{(\termlet{\termone}{x}{y}{\termtwo})\howe{\relone}\termseven}{\typeone}$} 
  \DisplayProof  
  $$
  \vspace{4pt}
  $$
  \AxiomC{$\tj{\contone}{\valone\howe{\relone}\valthree}{\typeone}$}
  \AxiomC{$\tj{\conttwo}{\valtwo\howe{\relone}\valfour}{\typetwo}$}
  \AxiomC{$\tj{\contone,\conttwo}{\termpair{\valthree}{\valfour}\relone\termone}{\typeone\otimes\typetwo}$}
  \TrinaryInfC{$\tj{\contone,\conttwo}{\termpair{\valone}{\valtwo}\howe{\relone}\termone}{\typeone\otimes \typetwo}$} 
  \DisplayProof  
  \vspace{4pt}
  $$
\end{minipage}}
\caption{The Howe's Rules for $\linL$.}\label{fig:howe}
\end{center}
\end{figure}
\begin{theorem}
In $\linL$, $\appsim$ is included in $\conpre$, thus $\appbis$ is included in $\conequ$.
\end{theorem}

\section{Injecting Probabilistic Choice}\label{sect:probchoice}
The expressive power of $\linL$ is rather limited, due to the presence
of linearity. Nevertheless, the calculus is complete for first-order
computations over the finite domain of boolean values, as discussed
previously. Rather than relaxing linearity, we now modify $\linL$ by
endowing it with a form or probabilistic choice, thus obtaining a new
linear $\lambda$-calculus, called $\linPL$, which is complete for
probabilistic circuits. We see $\linPL$ as an intermediate step
towards $\linQL$, a quantum $\lambda$-calculus we will analyze in the
following section.

The language of terms of $\linPL$ is the one of $\linL$ where,
however, there is one additional binary construct $\oplus$, to be
interpreted as probabilistic choice:
$\termone\bnf\termone\oplus\termone$.  The set $\types$ of types is
the same as the one of $\linL$.  An evaluation operation is introduced
as a relation $\bsred\subseteq \termsp{\linPL}{\emcon,A}\times
\distrsp{\linPL}{\typeone}$ between the sets of closed terms of type
$\typeone$ belonging to $\linPL$ and the one of subdistributions of
values of type $\typeone$ in $\linPL$.  The elements of
$\distrsp{\linPL}{\typeone}$ are actually subdistributions whose
support is some finite subset of the set of values
$\valuesp{\linPL}{\typeone}$, i.e., for each such $\distrone$, we have
$\distrone:\valuesp{\linPL}{A}\mapsto\mathbb{R}_{[0,1]}$ and
$\sum_{\valone\in\valuesp{\linPL}{\typeone}}\distrone(\valone)\leq 1$.
Whenever this does not cause ambiguity, subdistributions will be referred
to simply as distributions. In
Figure~\ref{fig:big-step-probabilistic-sem} a selection of
the rules for big-step semantics in $\linPL$ is given.
Expressions in the form $\{\valone_i^{\probone_i}\}_{i\in I}$ have the obvious meaning, namely
the distribution with support $\{\valone_i\}_{i\in I}$ which attributes probability
$\probone_i$ to each $\valone_i$.

As for the terms $\termone\in\termsp{\linPL}{A}$, the following lemma holds:
\begin{lemma}\label{lemma:unique-distribution}
  If $\tj{\emcon}{\termone}{\typeone}$, then there is a unique
  distribution $\distrone$ such that $\termone\bsred\distrone$.
\end{lemma}
Lemma~\ref{lemma:unique-distribution} only holds because the
$\lambda$-calculus we are working with is linear, and as a consequence
strongly normalising. If $\termone\bsred\distrone$, then the unique
$\distrone$ from Lemma~\ref{lemma:unique-distribution} is called the
\emph{semantics} of $e$ and is denoted simply as
$\semantic{\termone}$.
\begin{figure}
  \begin{center}
    \fbox{
      \begin{minipage}{.97\textwidth}
        \vspace{4pt}
        $$
        \AxiomC{$ $}\UnaryInfC{$\valone\Downarrow \{\valone^1\}$}\DisplayProof
        \qquad
        \AxiomC{$ $}\UnaryInfC{$\termdiv\Downarrow\emptyset$}\DisplayProof 
        \qquad
        \AxiomC {$\termone\Downarrow\distrone$}\AxiomC{$\termtwo\Downarrow\distrtwo$}
        \AxiomC{$\termfive\{\valtwo/x\}\Downarrow\distrthree_{\termabs{\varone}{\termfive},\valtwo} 
        $}		
        \TrinaryInfC{$\termone\termtwo\Downarrow \sum_{\termabs{\varone}{\termfive}\in\supp{\distrone}, \valtwo\in \supp{\distrtwo}} \distrone(\termabs{\varone}{\termfive})
          \distrtwo(\valtwo)\distrthree_{\termabs{\varone}{\termfive},\valtwo}$} 
        \DisplayProof
        $$
        \vspace{4pt}
        $$
        \AxiomC{$\termone\Downarrow \distrone$}
        \AxiomC{$\termtwo\Downarrow \distrtwo$} 
        \AxiomC{$\termthree\Downarrow\distrthree$}
        \TrinaryInfC{$\bigl(\termif{\termone}{\termtwo}{\termthree}\bigr) \Downarrow 
          \distrone(\termtt)\distrtwo+\distrone(\termff)\distrthree$} \DisplayProof
        \qquad
        \AxiomC{$\termone\Downarrow\distrone$}
        \AxiomC {$\termtwo\Downarrow\distrtwo$}	
        \BinaryInfC{$\termchoice{\termone}{\termtwo}\Downarrow \frac 12\distrone+\frac 12 \distrtwo$}
        \DisplayProof
        $$
        \vspace{4pt}
      \end{minipage}}
  \caption{Big-step Semantics of $\linPL$ --- Selection} \label{fig:big-step-probabilistic-sem}
  \end{center}
\end{figure}

Context equivalence and the context preorder are defined very
similarly to $\linL$, the only difference being the underlying notion
of observation, which in $\linL$ takes the form of \emph{convergence},
and in $\linPL$ becomes the \emph{probability} of convergence.

\subsection{Applicative Bisimilarity}\label{sect:probabilistic-applicative-similarity}
Would it be possible to define applicative bisimilarity for $\linPL$
similarly to what we have done for $\linL$? The first obstacle towards
this goal is the dynamics of $\linPL$, which is not deterministic but
rather probabilistic, and thus cannot fit into an LTS. In the
literature, however, various notions of probabilistic bisimulation
have been introduced, and it turns out that the earliest and simplest
one, due to Larsen and Skou~\cite{LarsenSkou-91}, is sufficient for
our purposes.

A \emph{labelled Markov chain} (LMC in the following) is a triple
$(\states,\labels,\transitionpm)$, where $\states$ and $\labels$ are
as in the definition of a LTS, while $\transitionpm$ is a
\emph{transition probability matrix}, i.e., a function from
$\states\times\labels\times\states$ to $\RR_{[0,1]}$ such that for
every $\statone$ and for every $\labelone$, it holds that
$\transitionpm(\statone,\labelone,\states)\leq 1$ (where the
expression $\transitionpm(\statone,\labelone,\setone)$ stands for
$\sum_{\stattwo\in\setone}\transitionpm(\statone,\labelone,\stattwo)$
whenever $\setone\subseteq\states$). Given such a LMC $\lmcone$, an
equivalence relation $\relone$ on $\states$ is said to be a
\emph{bisimulation} on $\lmcone$ iff whenever
$(\statone,\stattwo)\in\relone$, it holds that
$\transitionpm(\statone,\labelone,\econe)=\transitionpm(\stattwo,\labelone,\econe)$
for every equivalence class $\econe$ of $\states$ modulo $\relone$.  A
preorder $\relone$ on $\states$ is said to be a \emph{simulation} iff
for every subset $\setone$ of $\states$, it holds that
$\transitionpm(\statone,\labelone,\setone)\leq\transitionpm(\stattwo,\labelone,\relone(\setone))$.
With some efforts (see \cite{EV} for some more details) one can prove
that there exist largest bisimulation and simulation, that we continue
to call \emph{similarity} and \emph{bisimilarity}, respectively.
Probabilistic (bi)simulation, despite the endeavor required to define
it, preserves all fundamental properties of its deterministic sibling.
As an example, a symmetric probabilistic simulation is a
bisimulation. Moreover, bisimilarity is the intersection of similarity
and co-similarity.

Labelled Markov chains are exactly the objects we need when generalising
the construction $\ltsof{\linL}$ to $\linPL$. The LMC $\lmcof{\linPL}$,
indeed, is defined as the triple
$$
(\overline{\terms{\linPL}}\uplus\overline{\values{\linPL}},
\overline{\termsc{\linPL}}\uplus\overline{\values{\linPL}}\cup \{\eval,\termtt,\termff\}\cup(\types\uplus\types),\transpm{\linPL})
$$
where $\transpm{\linPL}$ is the function assuming the following values:
{\footnotesize
$$
\begin{array}{c}
\transpm{\linPL}((\dist{\termtt},\boolty), \termtt,
(\dist{\termtt},\boolty))=1;\qquad
\transpm{\linPL}((\dist{\termff},\boolty), \termff,
(\dist{\termff},\boolty))=1;\\
\transpm{\linPL}((\dist{\lambda\varone.\termone},\typeone\multimap\typetwo),(\valone,\typeone),(\subst{\termone}{\varone}{\valone},\typetwo))=1;\\
\transpm{\linPL}(( \widehat{\termpair{\valone}{\valtwo}},\typeone\otimes\typetwo ),
  (\termone,(\typeone, \typetwo,\typethree)),  (\termone\{\valone/\varone,\valtwo/\vartwo\} ,\typethree ))=1;\\
\transpm{\linPL}  ((\termone,\typeone),\typeone,(\termone,\typeone))=1
  \qquad
\transpm{\linPL} ((\dist{\valone},\typeone),\dist{\typeone},(\dist{\valone},\typeone))=1;\\
\transpm{\linPL}((\termone, \typeone),\eval, (\dist{\valone},\typeone))=\semantic{\termone}(\valone);
\end{array}
$$}

\noindent
and having value $0$ in all the other cases. It is easy to realise
that $\transpm{\linPL}$ can indeed be seen as the natural
generalisation of $\trans{\linL}$: on states in the form
$(\dist{\valone},\typeone)$, the function either returns $0$ or $1$,
while in correspondence to states like $(\termone,\typeone)$ and the
label $\eval$, it behaves in a genuinely probabilistic way.

As for $\linL$, simulation and bisimulation relations, and the largest
such relations, namely similarity and bisimilarity, can be given by
just instantiating the general scheme described above to the specific
LMC modeling terms of $\linPL$ and their dynamics. All these turn out
to be relations on \emph{closed} terms, but as for $\linL$, they can
be turned into proper typed relations just by the usual open extension.

The question now is: are the just introduced coinductive
methodologies sound with respect to context equivalence? And is it that
the proof of precongruence for similiarity from Section~\ref{subsect:appbis-congruence} can be
applied here? The answer is positive, but some effort is needed. More
specifically, one can proceed as in~\cite{DalLagoCrubille-14}, generalising
Howe's method to a probabilistic setting, which makes the Key Lemma
harder to prove. By the way, the set of Howe's rules are the same
as in $\linL$, except for a new one, namely
$$
\AxiomC{$\tj{\contone}{\termone\howe{\relone} \termfour}{\typeone}$}
\AxiomC{$\tj{\conttwo}{\termtwo\howe{\relone} \termfive}{\typeone}$}
\AxiomC{$\tj{\contone,\conttwo}{(\termchoice{\termfour}{\termfive})\relone\termseven}{\typeone}$}
\TrinaryInfC{$\tj{\contone,\conttwo}{(\termchoice{\termone}{\termtwo})\howe{\relone}
\termseven}{\typeone}$} 
\DisplayProof  
$$
Thus:
\begin{theorem}
In $\linPL$, $\appsim$ is included in $\conpre$, thus $\appbis$ is included in $\conequ$.
\end{theorem}
\section{On Quantum Data}\label{sect:quantum}
Linear $\lambda$-calculi with classical control and quantum data have
been introduced and studied both from an operational and from a
semantical point of
view~\cite{SelingerValiron-08,DengFeng-12}. Definitionally, they can
be thought of as $\lambda$-calculi in which ordinary, classic, terms
have access to a so-called quantum register, which models quantum
data.

A quantum register $\quregone$ on a finite set of quantum variables
$\quvars$ is mathematically described by an element of a
finite-dimensional Hilbert space whose computational basis
is the set $\sbsts{\quvars}$ of all maps 
from $\quvars$ to $\{\termtt,\termff\}$ (of which there are $2^{|\quvars|}$). 
Any element of this basis takes the form
$\quregket{\quvarone_1}{\qubitone_1}{\quvarone_2}{\qubitone_2}{\quvarone_n}{\qubitone_n}$,
where $\quvars=\{\quvarone_1,\ldots,\quvarone_n\}$ and
$\qubitone_1,\ldots,\qubitone_n\in\{\termtt, \termff\}$. 
Elements of this Hilbert space, called $\hilb{\quvars}$, are in the form
\begin{equation}\label{equ:qubit}
\quregone=\sum_{\sbstone\in\sbsts{\quvars}}\coeffone_\sbstone\ket{\sbstone},
\end{equation}
where the complex numbers $\coeffone_\sbstone\in\complexsp$ are the
so-called \emph{amplitudes}, and must satisfy the \emph{normalisation
  condition}
$\sum_{\sbstone\in\sbsts{\quvars}}|\coeffone_\sbstone|^2=1$. If
$\sbstone\in\sbsts{\quvars}$ and $\quvarone$ is a variable not
necessarily in $\quvars$, then
$\sbstone\{\quvarone\leftarrow\bitone\}$ stands for the substitution
which coincides with $\sbstone$ except on $\quvarone$ where it equals
$\bitone$.

The interaction of a quantum register with the outer environment can
create or destroy quantum bits increasing or decreasing the dimension
of $\quregone$.  This shaping of the quantum register is
mathematically described making use of the following operators:
\begin{varitemize}
\item 
  The probability operator $\probop{\bitone}{\quvarone}:
  \hilb{\quvars}\rightarrow\mathbb{R}_{[0,1]}$ gives
  the probability to obtain $\bitone\in \{\termtt, \termff\}$ as a result
  of the measurement of $\quvarone\in\quvars$ in the input register:  
  $$
  \probop{\bitone}{\quvarone}(\quregone)=\sum_{\sbstone(\quvarone)=\bitone}|\coeffone_{\sbstone}|^2.
  $$
\item
  If $\quvarone\in\quvars$, then the projection operator
  $\projop{\bitone}{\quvarone}:
  \hilb{\quvars}\rightarrow\hilb{\quvars-\{\quvarone\}}$ measures the
  variable $\quvarone$, stored in the input register, destroying the
  corresponding qubit. More precisely
  $\projop{\termtt}{\quvarone}(\quregone)$ and
  $\projop{\termff}{\quvarone}(\quregone)$ give as a result the
  quantum register configuration corresponding to a measure of the
  variable $\quvarone$, when the result of the variable measurement is
  $\termtt$ or $\termff$, respectively:
  $$
  \projop{\bitone}{\quvarone}(\quregone)=
  \left[\probop{\bitone}{\quvarone}(\quregone)\right]^{-\frac 12}
  \sum_{\sbstone\in\sbsts{\quvars-\{\quvarone\}}} \coeffone_{\sbstone\{\quvarone\leftarrow\bitone\}}\ket{\sbstone},
  $$
  where $\quregone$ is as in (\ref{equ:qubit}).
\item 
  If $\quvarone\not\in\quvars$, then the operator
  $\newop{\bitone}{\quvarone}:\hilb{\quvars}\rightarrow\hilb{\quvars\cup\{\quvarone\}}$
  creates a new qubit, accessible through the fresh variable name
  $\quvarone$, and increases the dimension of the quantum register by
  one .
\end{varitemize}
Qubits can not only be created and measured, but their value can also
be \emph{modified} by applying unitary operators to them.  Given any
such $n$-ary operator $\unopone$, and any sequence of distinct
variables $\quvarone_1,\ldots,\quvarone_n$ (where
$\quvarone_i\in\quvars$ for every $1\leq i\leq n$), one can build a
unitary operator $\unopone_{\quvarone_1,\ldots,\quvarone_n}$ on
$\hilb{\quvars}$.

\subsection{The Language}
We can obtain the quantum language $\linQL$ as an extension of basic
$\linL$. The grammar of $\linL$ is enhanced by adding the following
values:
$$
\termone\bnf\unopone(\valone)\midd \termmeas(\valone)\midd \termnew(\valone);
\qquad\qquad
\valone\bnf\quvarone;
$$ 
where $\quvarone$ ranges over an infinite set of quantum variables,
and $\unopone$ ranges over a finite set of unitary transformations.
The term $\termnew(\valone)$  acting on boolean constant, returns (a quantum
variable pointing to) a qubit of the same value, increasing this way
the dimension of the quantum register. The term
$\termmeas(\valone)$ measures a value of type qubit, therefore it decreases the dimension of the quantum register.

Typing terms in $\linQL$ does not require any particular efforts. The
class of types needs to be sligthly extended with a new base type for
qubits, called $\qubitty$, while contexts now give types not only to
classical variables, but also to quantum variables. The new typing
rules are in Figure~\ref{fig:quantum-typing-rules}.
\begin{figure}
  \begin{center}
    \fbox{
      \begin{minipage}{.97\textwidth}
        \vspace{4pt}
        $$
        \AxiomC{$\tj{\contone}{\valone}{\qubitty}$} 
        \UnaryInfC{$\tj{\contone}{\termmeas(\valone)}{\boolty}$} 
        \DisplayProof
        \qquad
        \AxiomC{$\tj{\contone}{\valone}{\boolty}$} \UnaryInfC{$\tj{\contone}{\termnew(\valone)}{\qubitty}$} 
        \DisplayProof
        $$
        \vspace{4pt}
        $$
        \AxiomC{$\tj{\contone}{\valone}{\qubitty^{\otimes n}}$} 
        \UnaryInfC{$\tj{\contone}{\termU(\valone)}{\qubitty^{\otimes n}}$} 
        \DisplayProof
        \qquad
        \AxiomC{}
        \UnaryInfC{$\tj{\quvarone:\qubitty}{\quvarone}{\qubitty}$} 
        \DisplayProof 
        $$
        \vspace{4pt}
    \end{minipage}}
  \caption{Typing rules in $\linQL.$}\label{fig:quantum-typing-rules}
  \end{center}
\end{figure}

The semantics of $\linQL$, on the other hand, cannot be specified
merely as a relation between terms, since terms only make sense
computationally if coupled with a quantum register, namely in a pair
in the form $\quclosure{\quregone}{\termone}$, which is called a \emph{quantum
  closure}. 
\begin{figure}
  \begin{center}
    \fbox{
      \begin{minipage}{.97\textwidth}
        \vspace{4pt}
        $$
        \AxiomC{$\qquad $}
        \UnaryInfC{$\quclosure{\quregone}{(\termabs{\varone}{\termone})\valone}\ssred \{\quclosure{\quregone}{\termone\{\valone/\varone\}}^1\}$}
        \DisplayProof 
        $$
        \vspace{4pt}
        $$
        \AxiomC{$\quclosure{\quregone}{\termone}\ssred\{\quclosure{\quregone_\idxone}{\termtwo_\idxone}^{\probone_\idxone}\}_{\idxone \in \Idxone} $} 
        \UnaryInfC{$\quclosure{\quregone}{\termone\termthree}\ssred\{\quclosure{\quregone_\idxone}{\termtwo_\idxone\termthree}^{\probone_\idxone}\}_{\idxone\in \Idxone}$} 
        \DisplayProof
        \qquad
        \AxiomC{$\quclosure{\quregone}{\termone}\ssred\{\quclosure{\quregone_\idxone}{\termtwo_\idxone}^{\probone_\idxone}\}_{\idxone \in \Idxone} $} 
        \UnaryInfC{$\quclosure{\quregone}{\valone\termone}\ssred\{\quclosure{\quregone_\idxone}{\valone\termtwo_\idxone}^{\probone_\idxone}\}_{\idxone\in \Idxone}$} 
        \DisplayProof
        $$
        \vspace{4pt}
        $$
        \AxiomC{}
        \UnaryInfC{$\quclosure{\quregone}{\termif{\termtt}{\termtwo}{\termthree}}\ssred \{\quclosure{\quregone}{\termtwo}^{1}\}$} 
        \DisplayProof
        \qquad
        \AxiomC{}
        \UnaryInfC{$\quclosure{\quregone}{\termif{\termff}{\termtwo}{\termthree}}\ssred \{\quclosure{\quregone}{\termthree}^{1}\}$} 
        \DisplayProof          
        $$
        \vspace{4pt}
        $$
        \AxiomC{$\quclosure{\quregone}{\termone}\ssred \{ \quclosure{\quregone_\idxone}{\termfour_\idxone}^{\probone_\idxone}\}_{ 
            \idxone \in \Idxone} $} 
        \UnaryInfC{$\quclosure{\quregone}{\termif{\termone}{\termtwo}{\termthree}}\ssred \{\quclosure{\quregone_\idxone}{\termif{\termfour_\idxone}{\termtwo}{\termthree}}^{\probone_\idxone}\}_{\idxone\in \Idxone}$} 
        \DisplayProof  
        $$
        \vspace{4pt}
        $$
        \AxiomC{$\stackrel{\qquad}{\qquad} $} 
        \UnaryInfC{$\quclosure{\quregone}{\termlet{\termpair{\valone}{\valtwo}}{\varone}{\vartwo}{\termtwo}}\ssred \{\quclosure{\quregone}{\termtwo\{\valone/\varone,\valtwo/\vartwo\}}^{1} \}$} 
        \DisplayProof 
        $$
        \vspace{4pt}
        $$
        \AxiomC{$\quclosure{\quregone}{\termone}\ssred \{ 
          \quclosure{\quregone_\idxone}{\termfour_\idxone}^{\probone_\idxone}\}_{\idxone\in\Idxone} $} 
        \UnaryInfC{$\quclosure{\quregone}{\termlet{\termone}{\varone}{\vartwo}{\termthree}}\ssred 
          \{\quclosure{\quregone_\idxone}{\termlet{ \termfour_\idxone} {\varone}{\vartwo}{\termthree}}^{\probone_\idxone}\}_{\idxone\in\Idxone}$} 
        \DisplayProof  
        $$
        \vspace{4pt}
        $$
        \AxiomC{$\stackrel{\quad}{\quad} $}
        \UnaryInfC{$\quclosure{\quregone}{\termmeas(\quvarone)} \ssred  \{ 
          \quclosure{\projop{\termff}{\quvarone}(\quregone)}{\termff}^{\probop{\termff}{\quvarone}(\quregone)}, 
          \quclosure{\projop{\termtt}{\quvarone}(\quregone)}{\termtt}^{\probop{\termtt}{\quvarone}(\quregone)}\}$}
        \DisplayProof 
        $$
        \vspace{4pt}
        $$
        \AxiomC{$\stackrel{\quad}{\quad} $}
        \UnaryInfC{$\quclosure{\quregone}{\termU\langle\quvarone_1,\ldots,\quvarone_n\rangle} \ssred  
          \{\quclosure{U_{\quvarone_1,\ldots,\quvarone_n}(\quregone)}{\langle\quvarone_1,\ldots,\quvarone_n\rangle}^1\}$}
        \DisplayProof
        $$
        \vspace{4pt}
        $$
        \AxiomC{$r\ \mbox{\footnotesize fresh variable}$}
        \UnaryInfC{$\quclosure{\quregone}{\termnew(\bitone)} \ssred  
          \{ \quclosure{\newop{\bitone}{\quvarone}(\quregone)}{\quvarone}^{1} \}$}
        \DisplayProof 
        \qquad
        \AxiomC{$\stackrel{\quad}{\quad} $}
        \UnaryInfC{$\quclosure{\quregone}{\termdiv} \ssred \emptyset$}
        \DisplayProof
        $$
        \vspace{4pt}
    \end{minipage}}
  \caption{Small-step Semantics of $\linQL$.}\label{fig:small-step-quantum-semantic}
  \end{center}
\end{figure}
Analogously to what has been made for $\linPL$, small step reduction
operator $\ssred$ and the big step evaluation operator $\bsred$ are
given as relations between the set of quantum closures and of quantum
closures distributions.  In figures
\ref{fig:small-step-quantum-semantic} and
\ref{fig:big-step-quantum-semantic} the small-step semantics and
big-step semantics for $\linQL$ are given.
\begin{figure}
  \begin{center}
    \fbox{
      \begin{minipage}{.97\textwidth}
        \vspace{4pt}
	$$
        \AxiomC{}
        \UnaryInfC{$\quclosure{\quregone}{\valone}\bsred \{\quclosure{\quregone}{\valone}^1\}$}
        \DisplayProof 
        \qquad
        \AxiomC{}
        \UnaryInfC{$\quclosure{\quregone}{\termdiv} \bsred \emptyset$}
        \DisplayProof
        \qquad
        \AxiomC{$r\ \mbox{\footnotesize fresh variable}$}
        \UnaryInfC{$\quclosure{\quregone}{\termnew(\bitone)}\bsred  \{\quclosure{\newop{\bitone}{r}(\quregone)}{\quvarone}^1\}$}
        \DisplayProof 
        $$
        \vspace{4pt}
        $$
        \AxiomC{} 
        \UnaryInfC{$\quclosure{\quregone}{U\langle\quvarone_1\ldots\quvarone_\idxfour\rangle}\bsred  \{\quclosure{U_{\quvarone_1,\ldots,\quvarone_\idxfour}(\quregone)}{\langle\quvarone_1,\ldots,\quvarone_\idxfour\rangle}^{1}\}$} 
        \DisplayProof  
        $$
        \vspace{4pt}
        $$
        \AxiomC{} 
        \UnaryInfC{$\quclosure{\quregone}{\termmeas(\quvarone)}\bsred 
          \{ \quclosure{\projop{\termff}{r}(\quregone)}{\termff}^{\probop{\termff}{r}(\quregone)}, \quclosure{\projop{\termtt}{r}(\quregone)}{\termtt}^{\probop{\termtt}{r}(\quregone)}\} $} 
        \DisplayProof  
        $$
        \vspace{4pt}
        $$
        \AxiomC{$\begin{array}{c}
          \quclosure{\quregone}{\termone} \bsred 
            \{\quclosure{\quregone_\idxone}{\termabs{\varone}{\termfour_\idxone}}^{\probone_\idxone}\}_{\idxone\in \Idxone}\\
          \quclosure{\quregone_\idxone}{\termtwo} \bsred 
            \{\quclosure{\quregone_{\idxone,\idxtwo}}{\termfive_{\idxone,\idxtwo}}^{\probtwo_{\idxone,\idxtwo}}\}_{\idxone,\idxtwo\in\Idxtwo}\\
          \quclosure{\quregone_{\idxone,\idxtwo}}{\termfour_\idxone\{\termfive_{\idxone,\idxtwo}/x\}} \bsred \distrone_{\idxone,\idxtwo}
          \end{array}$}
        \UnaryInfC{$ \quclosure{\quregone}{\termone\termtwo}\bsred \sum_{\idxone,\idxtwo}\probone_\idxone\cdot\probtwo_{\idxone,\idxtwo}\cdot\distrone_{\idxone,\idxtwo}$}
        \DisplayProof
        \qquad
        \AxiomC{$\begin{array}{c}
        \quclosure{\quregone}{\termone}\bsred \{ \quclosure{\quregone_\termff}{\termff}^{\probone_\termff},\quclosure{\quregone_\termtt}{\termtt}^{\probone_\termtt}\}\\  
        \quclosure{\quregone_\termff}{\termthree} \bsred\distrone\\
        \quclosure{\quregone_\termtt}{\termtwo} \bsred\distrtwo\\
        \end{array}$}
        \UnaryInfC{$\quclosure{\quregone}{\termif{\termone}{\termtwo}{\termthree}}\bsred \probone_\termff\distrone +
          \probone_\termtt\distrtwo$} 
        \DisplayProof  
        $$
        \vspace{4pt}
         $$
        \AxiomC{$\quclosure{\quregone}{\termone} \bsred  
                    \{ \quclosure{\quregone_\idxone}{\termpair{\valone_\idxone}{\valtwo_\idxone}}^{\probone_\idxone} \}_{\idxone\in \Idxone}$}
       	\AxiomC{$\quclosure{\quregone_{\idxone}}{\termtwo\{\valone_i/\varone,\valtwo_i/\vartwo\}  } \bsred
                    \distrone_{\idxone}$}
        \BinaryInfC{$ \quclosure{\quregone}{\termlet{\termone}{\varone}{\vartwo}{\termtwo}}
                  \bsred \sum_{\idxone}\probone_{\idxone}\cdot \distrone_{\idxone}   $}
        \DisplayProof
        $$
        \vspace{4pt}
     \end{minipage}}
  \caption{Big-step Semantics of $\linQL$.}\label{fig:big-step-quantum-semantic}
  \end{center}
\end{figure}
Quantum closures, however, are not what we want to compare, since 
what we want to be able to compare are \emph{terms}. Context equivalence,
in other words, continues to be a relation on terms, and can be specified
similarly to the probablistic case, following, e.g.~\cite{SelingerValiron-08}.
\subsection{Applicative Bisimilarity in $\linQL$}
Would it be possible to have a notion of bisimilarity for $\linQL$?
What is the underlying ``Markov Chain''? It turns out that LMCs as
introduced in Section~\ref{sect:probabilistic-applicative-similarity}
are sufficient, but we need to be careful. In particular, states of
the LMC are not terms, but quantum closures, of which there are in
principle nondenumerably many. However, since we are only interested
in quantum closures which can be obtained (in a finite number of
evaluation steps) from closures having an empty quantum register, this
is not a problem: we simply take states as \emph{those} closures,
which we dub \emph{constructible}. $\lmcof{\linQL}$ can be built
similarly to $\lmcof{\linPL}$, where (constructible) quantum closures
take the place of terms. The non zero elements of the function
$\transpm{\linQL}$ are defined as follows:

{\footnotesize
$$
\begin{array}{c}
\transpm{\linQL}  ((\widehat{\termtt},\boolty),\boolty,(\widehat{\termtt},\boolty))=1;\\
\transpm{\linQL}  ((\widehat{\termff},\boolty),\boolty,(\widehat{\termff},\boolty))=1;\\
\transpm{\linQL}((\quclosure{\quregone}{\dist{\quvarone}},\qubitty),
 (\quclosure{\quregthree}{\termone},\typeone),(\quclosure{\quregone\otimes\quregthree}{\subst{\termone}{\varone}{\quvarone}} ,\typeone))=1;\\ 
 \transpm{\linQL}((\quclosure{\quregone}{\dist{\termabs{\varone}{\termone}}},\typeone\multimap\typetwo),(\quclosure{\quregthree}{\hat{\valone}},\typeone),(\quclosure{\quregone\otimes\quregthree}{\subst{\termone}{\varone}{\valone}},\typetwo))=1;\\
\end{array}
$$
$$
\begin{array}{c}
 \transpm{\linQL}((\quclosure{\quregone}{\dist{\termpair{\valone}{\valtwo}}},\typeone\otimes\typetwo ),
  (\quclosure{\quregthree}{\termone},(\typeone, \typetwo, \typethree)),  (\quclosure{\quregone\otimes\quregthree}{\termone\{\valone/\varone,\valtwo/\vartwo\} },\typethree))=1;\\
\transpm{\linQL}((\quclosure{\quregone}{\termone},\typeone),
  \typeone,(\quclosure{\quregone}{\termone},\typeone))=1 \qquad  \transpm{\linQL}((\quclosure{\quregone}{\dist{\termone}},\typeone),
  \typeone,(\quclosure{\quregone}{\dist{\termone}},\typeone))=1;\\
  \transpm{\linQL}((\quclosure{\quregone}{\termone},\typeone),
  \eval,(\quclosure{\quregtwo}{\valone},\typeone))=\semantic{\quclosure{\quregone}{\termone}}\left(\quclosure{\quregtwo}{\valone}\right).
\end{array}
$$}

\noindent
Once we have a LMC, it is easy to 
apply the same definitional scheme we have seen for $\linPL$, and
obtain a notion of applicative (bi)similarity. Howe's method, in turn,
can be adapted to the calculus here, resulting in a proof of precongruence
and ultimately in the following:
\begin{theorem}
In $\linQL$, $\appsim$ is included in $\conpre$, thus $\appbis$ is included in $\conequ$.
\end{theorem}
More details on the proof of this can be found in~\cite{EV}.
\begin{example}
  An interesting pair of terms which can be proved bisimilar are
  the following two:
  $$
  \termone=\lambda\varone.\termif{(\termmeas\;\varone)}{\termff}{\termtt};
  \qquad
  \termtwo=\lambda\varone.\termmeas(X\;\varone);
  $$
  where $X$ is the unitary operator which flips the value of a qubit. 
  This is remarkable given, e.g. the ``non-local'' effects
  entanglement could cause.
\end{example}
\section{On Full-Abstraction}\label{sect:fullabst}
In the deterministic calculus $\linL$, bisimilarity not only is
\emph{included} into context equivalence, but \emph{coincides} with it
(and, analogously, similarity coincides with the context
preorder). This can be proved by observing that in $\ltsof{\linL}$,
bisimilarity coincides with trace equivalence, and each linear test,
i.e., each trace, can be implemented by a context. This result is not
surprising, and has already been obtained in similar settings
elsewhere~\cite{Bierman00}.

But how about $\linPL$ and $\linQL$? Actually, there is little hope to prove
full-abstraction between context equivalence and bisimilarity in a linear
setting if probabilistic choice is present. Indeed, as shown by van Breugel
et al.~\cite{BreugelMisloveOuaknineWorrell-05}, probabilistic bisimilarity can be characterised by a notion
of test equivalence where tests can be \emph{conjunctive}, i.e., they can be in the form
$\testone=\contest{\testtwo}{\testthree}$, and $\testone$ succeeds if both
$\testtwo$ and $\testthree$ succeeds. Implementing conjuctive tests, thus,
requires \emph{copying} the tested term, which is impossible in a linear
setting. Indeed, it is easy to find a counterexample to full-abstraction
already in $\linPL$. Consider the following two terms, both of which
can be given type $\boolty\multimap\boolty$ in $\linPL$:
$$
\termone=\abstr{\varone}{\weak{\varone}{\termtt\oplus\termff}};
\quad\qquad
\termtwo=(\abstr{\varone}{\weak{\varone}{\termtt}})\oplus(\abstr{\varone}{\weak{\varone}{\termff}}).
$$ The two terms are not bisimilar, simply because $\termtt$ and
$\termff$ are not bisimilar, and thus also
$\abstr{\varone}{\weak{\varone}{\termtt}}$ and
$\abstr{\varone}{\weak{\varone}{\termff}}$ cannot be
bisimilar. However, $\termone$ and $\termtwo$ can be proved to be
context equivalent: there is simply no way to discriminate between
them by way of a linear context (see~\cite{EV} for more details).

What one may hope to get is full-abstraction for extensions of the considered
calculi in which duplication is reintroduced, although in a controlled way.
This has been recently done in a probabilistic setting by Crubill\'e 
and the first author~\cite{DalLagoCrubille-14}, and is the topic of current investigations
by the authors for a non-strictly-linear extension of $\linQL$. 

\section{Conclusions}\label{sect:conclusions}
We show that Abramsky's applicative bisimulation can be adapted to linear
$\lambda$-calculi endowed with probabilistic choice and quantum data. The main
result is that in both cases, the obtained bisimilarity relation is a congruence,
thus included in context equivalence.

For the sake of simplicity, we have deliberately kept the considered
calculi as simple as possible. We believe, however, that many
extensions would be harmless. This includes, as an example,
generalising types to \emph{recursive} types which, although
infinitary in nature, can be dealt with very easily in a coinductive
setting. Adding a form of controlled duplication requires more care,
e.g. in presence of quantum data (which cannot be duplicated).

\bibliographystyle{plain}
\bibliography{biblio}
\end{document}